\documentclass[twocolumn,preprintnumbers,amsmath,amssymb]{revtex4}

\usepackage{graphicx}
\usepackage{dcolumn}
\usepackage{bm}

\begin{document}

\title{Why Don't We See the Hagedorn Mass  Spectrum  in the Experiments?  }

\author{K. A. Bugaev, V. K. Petrov  and G. M. Zinovjev}
\affiliation{Bogolyubov Institute for Theoretical Physics,
Kiev, Ukraine
}

\date{\today}
\begin{abstract}

The influence of medium dependent finite width of  the QGP bags on their  equation of state  is  analyzed on a basis of  an exactly solvable model 
with the general  mass-volume spectrum of these  bags. 
It is  arguing that the  consistent  statistical description of the  QGP bags is achieved 
for the width proportional to the square root of their volume. 
The model allows us  to estimate the minimal 
value  of the QGP bags' width  from the new lattice QCD data.
The large width of the QGP bags  not only explains 
the observed deficit in the number of  hadronic resonances compared to the  Hagedorn mass spectrum, but also clarifies the reason   why 
the heavy/ large  QGP bags   cannot be directly observed in experiments   as metastable  states in a hadronic phase.

\vspace*{0.25cm} 

\end{abstract}

\maketitle


\section{Introduction and formulating the problem}

{ Extensive experimental and theoretical searches for quark gluon plasma (QGP), i.e. the  deconfined phase of strongly interacting matter,  
became one of the focal point of the modern nuclear physics. 
The first signal for new physics at high energy densities 
was given by 
the statistical bootstrap model (SBM) \cite{Hagedorn:65}  
which  shows
that the exponentially increasing mass spectrum of hadrons, the Hagedorn spectrum,
could lead to new thermodynamics above the Hagedorn temperature $T_H$. 
Soon after this  it was found that  more sophisticated models like }
the dual resonance model (DRM) 
\cite{DRM, Miranski:73} (which originated the  string-like picture of hadrons) and the  bag model (which supposes the nontrivial vacuum structure) resemble the other features of SBM besides the asymptotic form of mass spectrum  \cite{MITBagM}. 
{Moreover, shortly after }
it has been realized that the Hagedorn temperature 
might be interpreted as the temperature of phase transition to the partonic degrees of freedom  \cite{Parisi:75}.
Henceforth these results initiated the extensive study of hadron 
thermodynamics within the model of a gas of bags (GBM) \cite{Kapusta:81}. 
The analytical solution
of GBM with a non-zero proper volume of hadronic bags (with the hard core repulsion)
allowed one to become aware of possible mechanism of deconfining phase transition from 
hadronic matter to the QGP (
{ represented}
 by an infinite bag containing 
free quarks and gluons)  \cite{Gorenstein:81}.
Amazingly,  up to now GBM remains one of the most 
efficient phenomenological instruments to successfully describe the bulk properties 
of hadron production in existing experimental data on relativistic heavy ion 
collisions and due to the simplicity of its foundations to easily incorporate newly 
discovered features of strongly interacting matter   \cite{HG}.
Apparently, the most recent 
attempts to update GBM bringing the contemporary knowledge of the  phase diagram
of  quantum  chromodynamics (QCD)
 \cite{Bugaev:05c,Bugaev:07,CGreiner:06}
are entirely based on the lattice approach to quantum chromodynamics (LQCD)
\cite{fodorkatz, karsch}. 

However, despite the considerable success of these models and their remarkable features all of them face two conceptual difficulties. The first one can be formulated by asking a very simple question: 'Why are the QGP bags never directly observed  in the experiments?' The routine argument applied to both high energy heavy ion and hadron collisions is that there exists  a phase transition and, hence, the huge energy gap separating the QGP bags from the ordinary (light) hadrons prevents the QGP co-existence at the hadron densities below the phase transition. The same line of  arguments is also valid  if the strong cross-over  exists. But on the other hand
in the laboratory experiments we are dealing with the finite systems and it is known  from the exact analytical solutions of the  constrained statistical multifragmentation model (SMM) \cite{Bugaev:04a}  and GBM \cite{Bugaev:05c} that there is a non-negligible probability to find the small and not too heavy QGP bags in thermally equilibrated finite systems  even in the cofined (hadronic) phase. 
Therefore,  for finite volume systems  created in  high energy nuclear or  elementary particle  collisions such QGP bags  could appear like any other  metastable states in statistical mechanics,
{since in this case  the statistical suppression is just a few orders  of magnitude and not of the order of   the Avogadro number.}
Moreover, at the pre-equilibrated stage of high energy collision nothing  can actually prevent their 
appearance. 
{ This very same  argumentation is true for the  strangelets 
\cite{Strangelets:A,Strangelets:B,Strangelets:C} whose intensive searches  
\cite{STRsearches:A, STRsearches:B,STRsearches:C}
in heavy ion collisions, in  many processes in the  universe and in the cosmic rays 
have not led to any convincing result. 
} 
Then, if  such QGP bags and { strangelets}  can be created there must be a reason which 
prevents their  direct  experimental  detection. 
{As we will show here there is an inherent property of the strongly 
interacting matter equation of state (EoS) 
which prevents their appearance 
inside of the  hadronic phase even in finite systems.  The same  property is also responsible for  the instability 
of  large or heavy  strangelets. 
}\\
\indent
The second conceptual  problem is
{ seen}  in   a huge deficit of the  number of observed  hadronic resonances \cite{Bron:04}  with masses above 2.5 GeV predicted by the SBM and used, so far,  by all other subsequent  models discussed above. 
{ Moreover, such a spectrum  has been derived on the basis of such profound models like
DRM \cite{DRM, Miranski:73},  bag model \cite{MITBagM} and GBM 
\cite{Kapusta:81, Goren:82}, but  the modern review of Particle Data Group 
contains  very few  heavier hadronic resonances 
comparing to the SBM expectations.
Furthermore, the best  description of particle yields observed in a very wide range of  
collision  energies of heavy ions   is  achieved 
by the statistical model which incorporates  all hadronic resonances not heavier than 2.3 GeV \cite{HG}.
Thus, it looks like  heavier hadronic species, except for the long living ones, are simply absent in 
the experiments \cite{Blaschke:03}. 

Hence,
} there is a paradox   situation with the Hagedorn mass  spectrum: it was predicted for heavy hadrons which nowadays  must be regarded as QGP bags, but  it can be experimentally  established up to hadronic masses of  about 2.3 GeV  \cite{Bron:04}.
Of course, one could  argue that heavy hadronic resonances cannot be established experimentally 
because both  their   large width  and  very large number of decay channels lead to  great  difficulties in their identification, but the point is that, despite the recent efforts  \cite{Blaschke:03},  the influence of   large  width of heavy 
resonances  on their EoS  properties and  the corresponding  experimental consequences 
were  not studied in full.

Therefore, here we would like 
{ to study  the role of 
finite  medium dependent
width of QGP bags, its influence onto  
the  EoS 
of system at  zero baryonic density and  show 
that  the novel physical  effect, the  {\it subthreshold suppression of the QGP bags}, generated  by  this finite width model (FWM)  resolves  both  the conceptual problems 
formulated above.  
As will be shown below  the FWM  also allows us to directly relate the obtained pressure of QGP bags at low temperatures to the LQCD pressure and to approximately estimate the width of these bags. 
}

\section{Main Ingredients of the  FWM}
The most convenient way to study the phase structure  of  any statistical  model similar to the SBM, GBM or  the QGP bags with surface tension model (QGBSTM) 
\cite{Bugaev:07}  implies to use the isobaric partition \cite{Gorenstein:81,Bugaev:07, Bugaev:04a} and find its rightmost singularities. Hence,  after the Laplace transform  the  FWM grand canonical  partition  $Z(V,T)$ generates the following 
isobaric partition:
\begin{eqnarray}\label{Zs}
\hspace*{-0.cm}\hat{Z}(s,T) \equiv \int\limits_0^{\infty}dV\exp(-sV)~Z(V,T) =\frac{1}{ [ s - F(s, T) ] } \,,
\end{eqnarray}
\noindent
where the function $F(s, T)$ contains the discrete $F_H$ and continuous $F_Q$ mass-volume spectrum 
of the bags 
\begin{eqnarray}
F(s,T)&\equiv& F_H(s,T)+F_Q(s,T) = \sum_{j=1}^n g_j e^{-v_js} \phi(T,m_j) \nonumber  \\
&+& \int\limits_{V_0}^{\infty}dv\hspace*{-0.1cm}\int\limits_{M_0}^{\infty}
 \hspace*{-0.1cm}dm~\rho(m,v)\exp(-sv)\phi(T,m)~.
 \label{FsHQ}
\end{eqnarray}
\noindent
The bag density
 of mass $m_k$, eigen volume $v_k$  and degeneracy factor $g_k$
is given by  $\phi_k(T) \equiv g_k ~ \phi(T,m_k) $  with 
\begin{eqnarray}
\phi_k(T)  & \equiv & \frac{g_k}{2\pi^2} \int\limits_0^{\infty}\hspace*{-0.1cm}p^2dp~
e^{\textstyle - \frac{(p^2~+~m_k^2)^{ \frac{1}{2} }}{T} } = \nonumber \\
& = &  g_k \frac{m_k^2T}{2\pi^2}~{ K}_2 {\textstyle \left( \frac{m_k}{T} \right) }\,.
\end{eqnarray}
\noindent
The mass-volume spectrum $\rho(m,v)$  generalizes   the exponential mass spectrum 
introduced by Hagedorn \cite{Hagedorn:65}.  As  in the GBM and QGBSTM,  the 
FWM bags 
are assumed to have the hard core repulsion of the Van der Waals type  generating  the suppression factor proportional to the  exponential of  bag 
proper  volume $\exp(-sv)$. 
The first term of Eq.~(\ref{FsHQ}), $F_H$, represents the contribution of a finite number of low-lying
hadron states up to mass $M_0 \approx 2 $ GeV \cite{Goren:82}. 
$F_H$ 
has no $s$-singularities at
any temperature $T$ and  generates  a simple pole  (\ref{Zs})
that describes a hadronic phase, whereas  
we will prove that 
the mass-volume spectrum of the bags $F_Q(s,T)$  
leads to  an essential  singularity $s_Q^* (T) \equiv p_Q(T)/T$ which defines  the QGP  pressure $p_Q(T)$  at zero baryonic densities 
\cite{Gorenstein:81,Goren:82, Bugaev:07}. 
Any   singularity  
$s^*$ of $\hat{Z}(s,T)$ (\ref{Zs}) 
is a solution of  Eq.    $s^*(T)~=~ F(s^*,T)$ \cite{Gorenstein:81,Bugaev:07}.

Here we use the simplest parameterization of  the  spectrum $\rho(m,v)$ to demonstrate the idea.
{Nevertheless, the requirements discussed in the introduction do not leave
us too much freedom to construct such a spectrum.
Thus, to have a firm bridge with the most general experimental and theoretical findings of 
particle phenomenology 
it is necessary to 
assume that the continuous  hadronic mass spectrum has a Hagedorn like form} 
\begin{eqnarray}\label{Rfwm}
 \rho (m,v) & = &   \frac{ \rho_1 (v)  ~N_{\Gamma}}{\Gamma (v) ~m^{a+\frac{3}{2} } }
 \exp{ \textstyle \left[ \frac{m}{T_H}   -   \frac{(m- B v)^2}{2 \Gamma^2 (v)}  \right]  } \,, \\ 
  \rho_1 (v) & = & f (T)\, v^{-b}~ \exp{\textstyle \left[  -  \frac{\sigma(T)}{T} \, v^{\varkappa}\right] }\,.
\label{R1fwm}
 \end{eqnarray}
{This spectrum has 
 the Gaussian attenuation  around the bag mass
$B v$ 
 determined by} 
 the volume dependent  Gaussian  width $\Gamma (v)$ or width hereafter. 
We will distinguish it from the true width defined as 
$\Gamma_R = \alpha \, \Gamma (v)$ ($\alpha \equiv 2 \sqrt{2 \ln 2}\,$).

{Usually  for narrow resonances  there used  two mass distributions, the Breit-Wigner 
and the Gaussian ones. As  will be shown later 
the Gaussian dependence is of a crucial importance  for the FWM because 
the Breit-Wigner attenuation leads to a divergency of the partition function. 
This is different from the early attempts to consider the  width of QGP bags in 
\cite{Blaschke:03}. 
}

The normalization factor in (\ref{Rfwm}) is defined to 
obey the condition
\begin{eqnarray}\label{Ng}
& N_{\Gamma}^{-1}~ = ~ \int\limits_{M_0}^{\infty}
 \hspace*{-0.1cm} \frac{dm}{\Gamma(v)}
    \exp{\textstyle \left[  -   \frac{(m- B v)^2}{2 \Gamma^2 (v)}  \right] } \,.
 \end{eqnarray}

It is important that the volume spectrum in  (\ref{R1fwm}) contains the surface free energy (${\varkappa} = 2/3$) with the $T$-dependent 
surface tension which is parameterized by 
$\sigma(T) = \sigma_0 \cdot
\left[ \frac{ T_{c}   - T }{T_{c}} \right]^{2k + 1} $  ($k =0, 1, 2,...$) \cite{Bugaev:07, Bugaev:04b},
where  $ \sigma_0 > 0 $ can be a smooth function of temperature. 
As shown in \cite{Bugaev:07}  such a  parametrization of  the bag surface tension 
is  necessary to generate the QCD tricritical endpoind.
For $T$ not above  the tricritical temperature $T_{c}$  this  form of $\sigma(T)$  is justified by the usual  cluster models 
like the Fisher droplet model  \cite{Fisher:67} and SMM \cite{Bondorf:95, Bugaev:00}, whereas 
the general   $T$ dependence    can be analytically derived from the surface partitions of the Hills and Dales model 
\cite{Bugaev:04b}. 
The important   consequences of such a
surface tension   and a discussion   of the curvature free energy 
absence in 
(\ref{R1fwm}) can be found in~\cite{Bugaev:07, Complement, HagedornTherm:06}.

An attempt of Ref. \cite{Goren:82} to derive  the bag pressure \cite{MITBagM}  within  the 
GBM  {is based on  a complicated  mathematical construct, but does not 
 explain any underlying  physical reason for the mass-volume 
spectrum of bags suggested in  \cite{Goren:82}.
In contrast to \cite{Goren:82}, 
the spectrum (\ref{Rfwm}) (and (\ref{R1fwm})) 
is  simple, but  general and adequate   for the medium dependence  of
both the width $\Gamma (v)$ and the bag's mass density $B$.}
It clearly reflects the fact 
that the QGP bags are similar to   the ordinary  quasiparticles with the medium dependent characteristics (life-time, most probable values of  mass and volume). 
Now we are ready to  
derive the  pressure  of an infinite bag
for two dependencies: the volume independent width $\Gamma(v) = \Gamma_0$ and 
the volume dependent width $\Gamma(v) = \Gamma_1 \equiv \gamma v^\frac{1}{2}$.

\section{The Continuous  FWM spectrum}
First we note that for large bag volumes ($v \gg M_0/B > 0$) the factor (\ref{Ng})  can be
found as  $N_\Gamma \approx 1/\sqrt{2 \pi} $.  Similarly, one can show that  for heavy free bags  ($m \gg B V_0$, $V_0 \approx 1$ fm$^3$ \cite{Goren:82},
{ignoring the  hard core repulsion and thermostat})
\begin{eqnarray}\label{Rm}
& \rho(m)  ~ \equiv   \int\limits_{V_0}^{\infty}\hspace*{-0.1cm} dv\,\rho(m,v) ~\approx ~
\frac{  \rho_1 (\frac{m}{B}) }{B ~m^{a+\frac{3}{2} } }
\exp{ \textstyle \left[ \frac{m}{T_H}     \right]  } \,,
\end{eqnarray}
\noindent
i.e. the  spectrum (\ref{Rfwm})
integrated over the bag  volume has a Hagedorn form modified by the surface free energy. 
It results from the fact that  for heavy bags the 
Gaussian  in (\ref{Rfwm}) acts as  a Dirac $\delta$-function for
either choice of $\Gamma_0$ or $\Gamma_1$. 
Thus, the Hagedorn form of  (\ref{Rm}) has a clear physical meaning and gives an additional argument in favor of the FWM. Also it gives an upper bound for the 
volume dependence of $\Gamma(v)$: the Hagedorn-like mass spectrum (\ref{Rm}) can be derived, if for large $v$ the width  $\Gamma$ increases  slower than $v^{(1 - \varkappa/2)}= v^{2/3}$. 

Similarly to (\ref{Rm}), one can estimate the width of heavy free bags  averaged over their  volumes and get  $ \overline{\Gamma(v) } \approx  \Gamma(m/B) $.
Thus, 
with choosing   $\Gamma(v) = \Gamma_1(v)$ the mass spectrum of heavy free QGP bags 
must be the Hagedorn-like one and  heavy resonances 
 develop 
the large  mean width $ \Gamma_1(m/B) = \gamma \sqrt{m/B}$. Hence,  they 
are hard to be observed. 
{Applying these arguments to the strangelets,
we conclude  that, if their mean volume is a few cubic fermis or larger, they  should survive for a  very short time,
which is in line with the results of \cite{Strangelets:06}.

Note also that such a mean width is essentially different from both the linear mass dependence of string models  \cite{StringW} and from an  exponential  form  of the nonlocal field theoretical models \cite{NLFTM}.}

Next we  calculate  $F_Q(s,T)$ (\ref{FsHQ}) for the  spectrum (\ref{Rfwm}) performing the mass integration. There are two distinct 
options 
 depending on the sign of the most probable mass: 
\begin{eqnarray}\label{Mprob}
& \langle m \rangle ~ \equiv ~  B v + \Gamma^2 (v) \beta\,,\quad {\rm with} 
\quad \beta \equiv  T_H^{-1} - T^{-1} \,. 
\end{eqnarray}
If {\boldmath 
$ \langle m \rangle > 0$} for $v \gg V_0$,  one can use the saddle point method
for mass integration to  find  the function~$F_Q (s,T)$
\begin{eqnarray}\label{FQposM}
&  F_Q^+ (s,T)   \approx \left[  \frac{T}{2\pi} \right]^{\frac{3}{2} }
\int\limits_{V_0}^{\infty}dv ~ \frac{ \rho_1(v) }{\langle m \rangle^a} ~\exp{\textstyle \left[  \frac{(p^+  - sT )v}{T}  \right]} \, 
\end{eqnarray}
\noindent
and the pressure of large  bags 
$p^+ \equiv T \left[ \beta B + \frac{\Gamma^2 (v)}{2 v} \beta^2 \right]$.
To get  (\ref{FQposM}) one has to employ  in (\ref{FsHQ}) an asymptotics  of the $K_2$-function  $\phi(T,m)\simeq (mT/2\pi)^{3/2}\exp(-m/T)$ for $m\gg~T$, 
{collect all $m$-dependent terms 
in exponential, get a full square for $(m -  \langle m \rangle)$ 
} and 
perform 
the Gaussian integration.

Since for $s  <  s_Q^*(T) \equiv p^+(v\rightarrow \infty)/T $ the integral  (\ref{FQposM}) diverges on its upper limit, 
the  partition (\ref{Zs}) has  an essential singularity corresponding to 
the QGP pressure of  an  infinite   bag. 
It allows one  to conclude 
the width  $\Gamma$ cannot increase  faster than $v^{1/2}$ for $v\rightarrow \infty$, otherwise $p^+(v\rightarrow \infty) \rightarrow \infty $ and  $F_Q^+ (s,T)$ diverges for any $s$.
Thus,  for {\boldmath $ \langle m \rangle > 0$} the phase structure of the FWM  with  $\Gamma (v) \neq 0$ is similar to the QGBSTM \cite{Bugaev:07}.

The bag spectrum  $F_Q^+ (s,T)$ (\ref{FQposM}) is of general nature  and, in contrast  to
the suggestion  of \cite{Goren:82},  has a transparent  physical origin. One can also see that  two general  sources  of the   bag pressure 
\begin{eqnarray}\label{PposM}
&  
p^+  =  \frac{T}{v} \left[ \beta\, \langle m \rangle - \frac{1}{2}\,\Gamma^2 (v) \beta^2 \right]
\end{eqnarray}
\noindent
are the bag most  probable   mass  and its width. 
Different $T$ dependent functions $\langle m \rangle$ and 
$\Gamma^2 (v)$ 
lead to 
different 
EoS. 

{If  instead of  the Gaussian width parametrization  in
(\ref{Rfwm}) we used  the Breit-Wigner one, then we would not be able to derive 
the continuous spectrum $F_Q^+ (s,T)$ (\ref{FQposM})  and the corresponding bag pressure for any  nonvanishing bag  width $\Gamma (v)$. Indeed,  for $T > T_H$ the mass integrals in  $F_Q (s,T)$
would diverge like in SBM, unless the Breit-Wigner mass attenuation has a zero width or an exponentially increasing width 
$\Gamma \sim \exp [m /T_H ]$   \cite{Blaschke:03}.
The former does not resolve the both of the GBM conceptual problems,
whereas the latter corresponds to a very specific  ansatz  for the resonance  width which is in contradiction with the FWM assumptions. } 

It is  possible to use the spectrum (\ref{FQposM}) not only for infinite system volume but for 
finite volumes $V \gg V_0$ as well. In this case the upper limit of integration should be replaced by finite $V$ 
(see Ref. \cite{Bugaev:05c} for details).  It changes  the singularities of partition 
 (\ref{Zs}) to a set of simple poles  $ s_n^*(T)$ in the complex $s$-plane which are  defined by the same equation as for 
 $V \rightarrow \infty$.  Similarly to the finite $V$ solution of the GBM 
\cite{Bugaev:05c},  it can be shown that for finite $T$ the FWM  simple poles may have  a small positive or even negative real part which would lead to a non-negligible contribution of the QGP bags into the  spectrum  $F(s,T)$  (\ref{FsHQ}).
Thus,
if the spectrum (\ref{FQposM})  was the only volume spectrum of the QGP bags, then there would exist the non-negligible probability of finding   heavy QGP bags ($m \gg M_0$)  in finite systems  at 
 $T \ll T_H$.  
Therefore, using the results of   the finite volume GBM and SMM,  we  conclude that the spectrum 
(\ref{FQposM}) itself 
cannot  explain  the absence of  the QGP bags at  $T \ll T_H$  and, hence, an alternative explanation of this fact is required. 

Such an explanation corresponds to the values  {\boldmath$ \langle m \rangle \le 0 $} for $v \gg V_0$.
From (\ref{Mprob}) one can see that  
for the volume dependent width $\Gamma (v) = \Gamma_1 (v) $ the most probable mass $ \langle m \rangle $ inevitably becomes negative at low $T$, if $0 < B < \infty$. 
{Using the asymptotics of the $K_2$-function for large and small values of $\frac{m}{T}$ one can show that 
at low $T$  the maximum of the  Gaussian mass distribution is located  at 
$ \langle m \rangle \le 0$. 
Hence only the tail  of  
the  Gaussian mass distribution  close to $M_0$  contributes to $F_Q(s,T)$. 
By the steepest descent method and with the $K_2$-asymptotic form   for $M_0 T^{-1} \gg 1$ one gets}
\begin{eqnarray}\label{FQnegM}
\hspace*{-0.5cm}F_Q^-(s,T) \hspace*{-0.05cm} \approx  \hspace*{-0.075cm} \left[  \frac{T}{2\pi} \right]^{\hspace*{-0.05cm}\frac{3}{2} } 
\hspace*{-0.15cm}
\int\limits_{V_0}^{\infty} \hspace*{-0.15cm} dv  \frac{ \rho_1(v) N_{\Gamma}\, \Gamma (v)\, \exp{\textstyle \left[  \frac{(p^-  - sT )v}{T}  \right]}
}{M_0^a\, [M_0 - \langle m \rangle  + a \, \Gamma^{2} (v)/ M_0 ]} \hspace*{-0.3cm}
\end{eqnarray}
\noindent
with the analytic form for the QGP bag pressure  
\begin{eqnarray}\label{pnegM}
p^-\big|_{v \gg V_0}  = {\textstyle  \frac{T}{v} \left[  \beta M_0 -  \frac{(M_0 - Bv)^2}{2\, \Gamma^{2} (v)}  
 \right] }\,.
\end{eqnarray}
\noindent
We would like to stress   the last result requires $B>~0$ and  cannot be generated  by  a weaker $v$-dependence  than  $\Gamma(v) = \Gamma_1(v)$. 
Indeed, if $B<0$, then the normalization factor (\ref{Ng}) would not be $1/\sqrt{2 \pi}$, but changes to 
$N_{\Gamma} \approx   [M_0 - \langle m \rangle]\, \Gamma^{-1} (v) \exp{\textstyle 
\left[    \frac{(M_0 - Bv)^2}{2\, \Gamma^{2} (v)}  \right]} $ and, thus,  it would cancel 
the leading  term in  pressure (\ref{pnegM}). Note that  the  inequality 
{\boldmath$ \langle m \rangle \le 0 $} for all $v \gg V_0$ with  $B > 0$ and 
finite $p^-(v \rightarrow \infty)$ is valid 
for  $\Gamma(v) = \Gamma_1(v)$ only.
{ 
The negative values of  $ \langle m \rangle $ serve as an indicator of a different  physical
situation comparing to $ \langle m \rangle > 0$, but    have no physical meaning since 
$ \langle m \rangle \le 0 $ does not enter   the main physical observable  $p^- $.

Also it is necessary to point out  that the only width $\Gamma(v) = \Gamma_1(v)$ 
does not lead to any difficulties  with the pressure of the bag in thermodynamic limit. 
This is clearly seen from Eqs. (\ref{PposM}) and (\ref{pnegM}) since the multiplier 
$\Gamma^2(v)$ stands in the numerator  of the pressure (\ref{PposM}), whereas 
in the pressure (\ref{pnegM}) it appears in the denominator.  Thus, if one chooses 
the different $v$-dependence  for the  width, then either $p^+$ or $p^-$  would 
diverge for the bag of  infinite  size. 
}

The new outcome of this case with $B>0$ is that for $T < T_H$ the spectrum 
(\ref{FQnegM}) contains the lightest QGP bags having the  smallest volume since 
every term in the pressure (\ref{pnegM}) is negative.  The finite volume of the system is no longer  important   because only  the  smallest bags survive in (\ref{FQnegM}).
Moreover, if such bags are created, they would have the  masses  of about  $M_0$ and
the widths of  about $\Gamma_1(V_0)$, and, hence, they would not be distinguishable 
from the usual low-mass hadrons. 
Thus, the option  {\boldmath$ \langle m \rangle \le 0 $} with 
$B>0$ leads to the {\it subthreshold suppression of the QGP bags} at low temperatures,
since their most  likely  mass is below the mass threshold  $M_0$ of the spectrum $F_Q(s,T)$.  Note that such an effect cannot be derived within  any of  the GBM-kind models  proposed earlier.

\section{Comparison with lattice QCD }

The obtained results give us a unique opportunity to make a bridge between the
particle phenomenology, some experimental facts and LQCD. 
Let us consider 
several examples of the  
QGP  EoS  and relate them  to the above results. First, we study the possibility to get  the bag model pressure 
 $p_{bag} \equiv \sigma T^4 - B_{bag} $  \cite{MITBagM}  by  the  stable QGP bags, i.e. 
$\Gamma (v) \equiv 0$. Equating the pressures $p^+$ and $p_{bag}$, one finds that 
$T_H$
must be related to a bag constant  as
$B_{bag} \equiv \sigma T^4_H$.  Then the mass density of such bags 
$\frac{\langle m \rangle}{v} \equiv  B  =  \sigma T_H (T  + T_H)(T^2 + T_H^2)$ is
always positive. Thus, the bag model  EoS  can be easily obtained by 
the FWM approach, but, as  discussed earlier, such bags should have been  observed. \\ \indent
Secondly,  we consider the stable bags, $\Gamma (v) \equiv 0$, but without the Hagedorn 
spectrum, i.e. $T_H \rightarrow \infty$. 
Matching $p^+ = - B $ and $p_{bag}$,  we find that at low temperatures 
the bag mass density  $\frac{\langle m \rangle}{v} = B$ is positive, whereas for 
high $T$ the mass density cannot be positive and, hence,  
one cannot reproduce $p_{bag}$ as $B \le 0$ and  the resulting pressure is not $p^-$ (\ref{pnegM}),
but a zero, as seen from (\ref{FQnegM}), (\ref{pnegM}) and   $N_{\Gamma}$ expression for the limit $\Gamma (v) \rightarrow 0$. 
One can try to reproduce $p_{bag}$ with the finite $T$ dependent  width $\Gamma (v) = 2 \, \sigma T^5 v$ for  $T_H \rightarrow \infty$. Then one can get  $p_{bag}$ from 
$p^+$, but only for low temperatures obeying the inequality 
$\frac{\langle m \rangle}{v} = B_{bag} - 2\, \sigma T^4 > 0$. Thus,  these  two 
examples  teach  us that without the Hagedorn mass spectrum one cannot  get 
the  bag model pressure. 

{
The FWM is a phenomenological model in which there exist two independent functions, $B$ and 
$\gamma$, 
parameterizing the QGP bag pressure and hence it requires additional information as an input. 
However, the FWM  provides us with some general results. 
Thus,  one can get  the general conclusions on   the temperature dependence of the QGP pressure in the limit 
$T \rightarrow 0$. For nonvanishing width coefficient  $\gamma_0 \equiv \gamma (T=0) > 0$  there exist  two possibilities. The first one corresponds to finite  $B_0 \equiv B (T=0) >0$ value. 
Then from (\ref{pnegM}) one concludes that  in the limit  $T \rightarrow 0$ the QGP pressure 
linearly depends on temperature $p^-(v \rightarrow \infty) \rightarrow - T\frac{ B_0^2}{2\, \gamma_0^2 } $. 
The second possibility corresponds to the divergent behavior of $B \rightarrow \frac{g_0}{T^D}$ (with $D > 0$)
provided that $\frac{\langle m  \rangle}{v} < 0$ for $ v \rightarrow \infty$.
The latter requires that $D \le 1$ for finite $\gamma_0$. 
In this case at  $T \rightarrow 0$ the QGP pressure should  behave  as 
$p^-(v \rightarrow \infty) \rightarrow - \frac{ g_0^2}{2\, \gamma_0^2 } \, T^{1 - 2D} $. 
Note that either of these possibilities  is a manifestation of the nonperturbative effect since in the limit $\gamma = 0$
they cannot be obtained. 

\begin{figure}[ht]

\includegraphics[width=7.7cm,height=5.72cm]{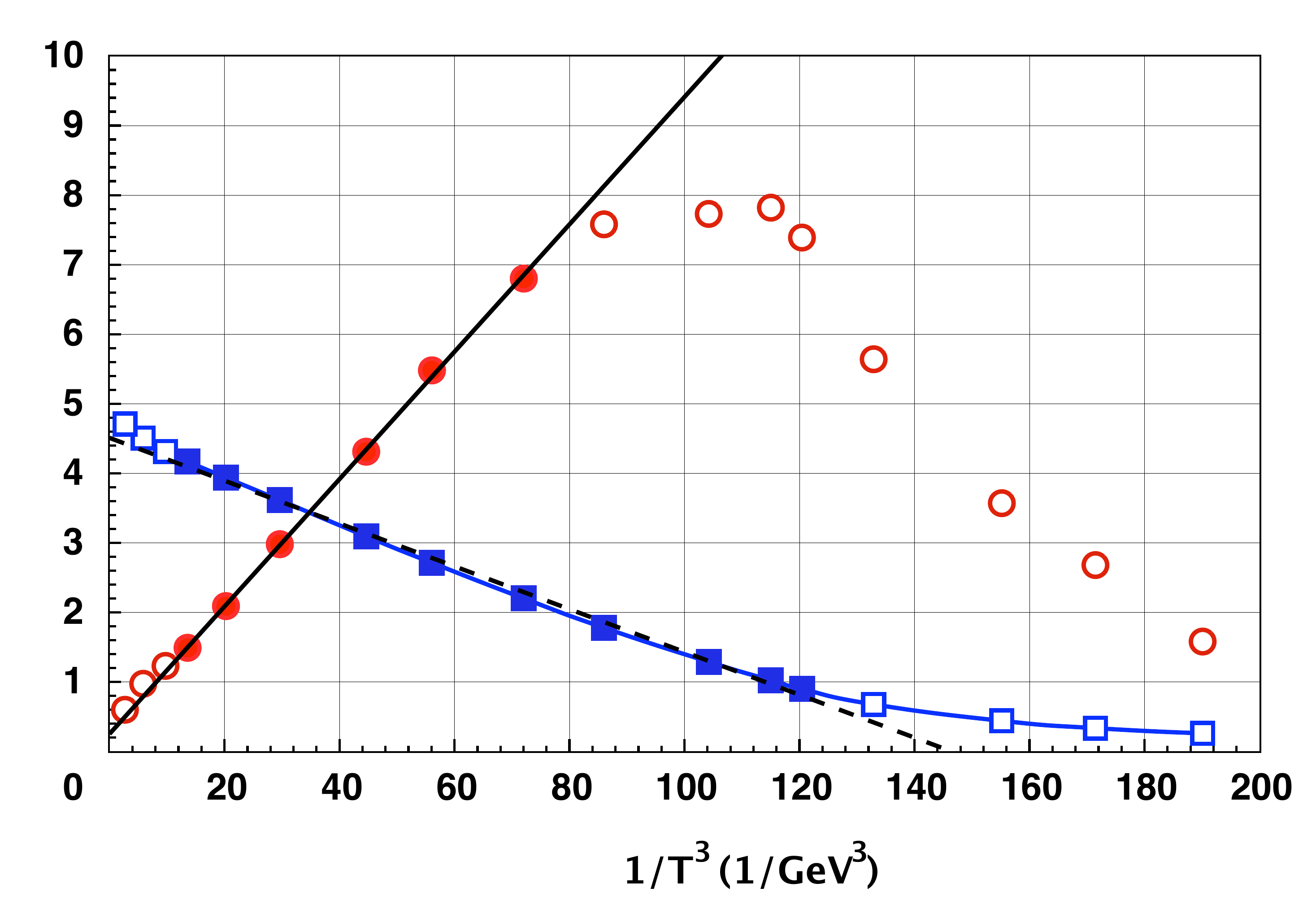}

\caption{
LQCD data for trace anomaly 
(circles) and  pressure  per $T^4$  (squares)  
as the functions of $T^{-3}$. Straight lines represent the fit of the filled symbols. 
See details in the text. The curve connecting the squares is to guide the eyes. 
}
\label{fig1}
\end{figure}

Long ago it was found    \cite{GorMog} that
the LQCD data \cite{LQCD:1,LQCD:2} exhibit the first of these possibilities. 
The corresponding EoS of the QGP has an additional  linear temperature dependence,
$p_a = \sigma T^4 - A_1 T + A_0 $ ($ A_1 > 0$, $ A_0 
\ge  0$).
However, the recent analysis \cite{Rob:07} of more fresh LQCD data \cite{LQCD:96} demonstrated not a linear, but the quadratic $T$-dependence of the trace anomaly and pressure
in the range  of temperatures between about $1.1 T_c$ and  $4 T_c$. 
Therefore, to clarify the question of an additional $T$-dependence of the LQCD pressure we performed the analysis of the newest LQCD data  found for the almost physical quark masses
\cite{LQCD:3}. 
The fitting of  $p \, T^{-4}$ as function of $T^{-3}$ shown in Fig. 1 by dashed line clearly demonstrates 
the linear  $T^{-3}$ dependence  $p \, T^{-4} = a_0 + a_1\, T^{-3} $ with 
$a_0 \approx  4.5094  $
and  $a_1 \approx - 0.0304$ GeV$^3$ for  ten data points in the range $T \in [202.5; 419.09]$ MeV. 

The linear $T$ dependence of pressure is born in the behavior of the trace anomaly $\delta = (\varepsilon - 3 \,p) T^{-4} $   (here $\varepsilon$ denotes the energy density). Indeed, plotting $\delta$ as the function of $T^{-3}$ (see circles in Fig. 1)  we found three different types of behavior. 
As one can see from Fig. 1  up to  $T^{-3} \approx 72.056$ GeV$^{-3}$ (for $T \ge  240.31$ MeV) the function
$\delta$ grows nearly linearly, and for  $T^{-3} \ge 120.43$ GeV$^{-3}$ (or $T \le 202.5$ MeV)
it decreases nearly linearly, whereas in between these values of $T^{-3}$ the function
$\delta  $ remains almost constant. 
The analysis shows that six LQCD data points of the function $\delta$ which belong  to  the range $T^{-3} \in [13.585; 72.056]$ GeV$^{-3}$   are, indeed, described by 
\begin{eqnarray}\label{TraceAN}
\delta = \tilde{a}_0 + \tilde{a}_1\, T^{-3} 
\end{eqnarray}
\noindent
with $\tilde{a}_0 \approx 0.2514 $
and  $\tilde{a}_1 \approx 0.0916$ GeV$^3$ and 
 $\chi^2/d.o.f. \approx 0.063$, i.e. with extremely 
high accuracy.  
The linear  $T^{-3}$-dependence of   $p \, T^{-4}$  is observed in a slightly  wider range of  $T^{-3}$
because of the approximately constant behavior of  the $\delta$ function at the moderate  values of  $T^{-3}$, but  with  lower  quality of the fit which,
however, is comparable with that one of Ref.  \cite{GorMog}.

The reason for  lower quality   of the pressure fit    can be seen  from  its relation to  the lattice  trace anomaly
\begin{eqnarray}\label{pT4}
\frac{p_{fit}}{T^4} - \frac{p_{0}}{T_0^4} &= &  \int\limits_{T_0}^T \hspace*{-0.0cm} d\,T \, \frac{\delta}{T}  \nonumber \\
&= & \tilde{a}_0 \ln\left[ \frac{T}{T_0} \right] - \frac{\tilde{a}_1}{3}\, \left[ T^{-3} -  T_0^{-3} \right]\,,  
\end{eqnarray}
%
\noindent
and, hence, one gets   $a_1 = - \frac{\tilde{a}_1}{3} $,  which is well supported by the LQCD data.
The last equality in (\ref{pT4}) is obtained from the linear fit of  $\delta$ and, hence,
$T_0$ and $p_0 \equiv p_{fit} (T_0)$ are the constants of integration. 

Eq. (\ref{pT4}) shows that for the temperatures between $240.31$ and  $419.09$ MeV
the LQCD pressure \cite{LQCD:3}  does not have a constant term, i.e. $A_0 =0$ for 
$p_a$, but
there exist higher order corrections ($T^5$ and higher) to  pressure.
They  are very small in this range of temperatures since $\tilde{a}_0 \ll 1$, but, in principle,  can be taken into account to improve the quality of 
the linear fit of  $p \, T^{-4}$ function suggested  in \cite{GorMog}. 
However, our main point is that either rough or refined analysis of the  modern LQCD data 
strongly suggests an existence of the linear $T$-dependent term in the LQCD pressure 
for $T \in [ 240.31; 419.09]$ MeV.

Since neither the nonrelativistic hadron gas with the hard core repulsion represented by  $F_H(s,T)$ in (\ref{FsHQ}) nor its relativistic analog 
analyzed in \cite{Bugaev:08NPA} can generate the linear $T$ dependence 
of pressure, it is possible that such a dependence is an inherent property of 
the LQCD data.  Assuming this, one obtains  that at low $T$ 
the LQCD pressure of the QGP phase  should behave as $p_{QGP} (T \rightarrow 0) \rightarrow - |a_1|\,  T$. 

Furthermore, the linear $T$-dependence of  the LQCD  pressure  (\ref{pT4}) evidences  that the FWM pressure at low temperatures  (\ref{pnegM}) correctly catches 
the nonperturbative features of the QGP EoS and, hence,  is one of the strongest arguments in favor of the FWM.

Moreover, such a behavior  of  the QGP   pressure  at low $T$ 
allows us to roughly  estimate 
the width  $\Gamma_1 (V_0)$. 
The FWM pressure depends on two functions and, hence,  in order 
to find them it is necessary to  know the form of the QGP pressure   in the domain
of hadronic phase.  Unfortunately the present  LQCD data do not provide 
us with such a detailed information and, thus, some additional 
assumptions are unavoidable. 
Neglecting  a very small  logarithmic term in  (\ref{pT4}) we obtain  the  pressure 
$p_{fit} = a_0 T^4 - |a_1| T$ found above from the fitting the LQCD data  \cite{LQCD:3}
 with   $a_0 =   p_0 T_0^{-4} +  |a_1| T_0^{-3} \approx  4.5094$.
It is convenient to fix $T_0 = T_H$ for which $p_0 \equiv p_{fit} (T_0) = 0$ as required by 
pressure $p^+$.  Then one determines $T_H  = \left[ \frac{|a_1|}{a_0} \right]^{\frac{1}{3}} \approx 188.91 $ MeV. 
Matching $p_{fit}$  with $p^- (v \rightarrow \infty) = - T \frac{B^2}{2\, \gamma^2}   $ for $T 
\rightarrow 0$, 
we can determine  $B^2/\gamma^2 $ ratio at this  temperature $B^2/\gamma^2 \equiv 
B^2_0/\gamma^2_0 = 2 | a_1|$. 
On the other hand, equating  $p_{fit}$ and $p^+$, one obtains the width coefficient  for $T \ge c_\pm\, T_H$ (here $c_\pm < 1$, for its definition see below)
\begin{eqnarray}\label{gammaI}
\gamma^2 = 2 \, \beta^{-1} \left[  a_0  T_H T (T^2 + TT_H + T_H^2) - B(T) \right] \,. 
\end{eqnarray}
To have 
a positive finite width in the whole  vicinity of $T_H$, it is necessary that  $(T-T_H)$ is a divisor 
of the difference staying  in the square brackets in (\ref{gammaI}). The simplest possibility for this is to suppose that 
\begin{eqnarray}\label{BI}
B(T) =  a_0   T_H^2  (T^2 + TT_H + T_H^2) 
\end{eqnarray}

\vspace*{-0.05cm}
\noindent
for any $T$.  
Evidently, $B(T)$ in (\ref{BI}) is positive and does not vanish at $T= 0$. 
In addition to a simplicity  the advantage of such a choice is that (\ref{BI}) does not require 
any new constant or  any new function  which are not involved
in (\ref{gammaI}).

Then equating $p_{fit}$ and $p^- (v \rightarrow \infty) = - T \frac{B^2}{2\, \gamma^2}   $, it 
is possible to completely determine  $\gamma^2 = \frac{B^2(T)}{2 a_0 (T_H^3 - T^3)}$  for  $T \le c_\pm  T_H$, if  (\ref{BI}) defines $B(T)$. 
For $T=0$   this   gives us 
$ \gamma_0^2 = B_0^2 / (2 a_0 T_H^3)  =  B_0^2 / (2 |a_1|) = T_H B_0 / 2 = a_0  T_H^5 /2$.
 Similarly,  from   (\ref{gammaI})  and (\ref{BI}) 
one obtains  $\gamma^2 = 2 \, T B(T) $ for $T \ge c_\pm  T_H$. 
The switch temperature $T_\pm = c_\pm  T_H = T_H/2$  can be found by equating 
$\gamma^2$ values obtained for  temperatures  below and above  $T_\pm$.

From these results we  find
 the true width for the SU(3) color group with three quark flavors  (see model C in the Table I)
to be 
$\Gamma_R (V_0, T=0) \approx 1.304\,  V_0^{\frac{1}{2}} \, T_c^{\frac{5}{2}} \alpha \approx 596 $ MeV and
$\Gamma_R (V_0, T=T_H)   = \sqrt{12}\,\Gamma_R (V_0, T=0) \approx 2066$ MeV. 
Such  estimates clearly demonstrate us  that there is no way to detect the decays of such
shortly living  QGP bags. 
The sensitivity of these results to  $T_c$ value and the number of elementary degrees of freedom of the LQCD  is given  in the Table I.  
As one can see from  the Table I the minimal width of the QGP bags  found for the same value of 
the transition temperature $T_c$  practically 
does not depend on the number of elementary  degrees of freedom  of the QGP. 
Such a finding not  only  supports our main conclusion for the short life time of the QGP bags, but   is a good argument in favor of the ansatz  (\ref{BI}).

A more detailed analysis of the LQCD pressure (\ref{pT4}) is performed in \cite{Reggeons:08}.
There it is found that within a few percent the above results remain valid, if one accounts for the
logarithmic  term in (\ref{pT4}).

}

\vspace*{0.3cm}


\noindent{\footnotesize{
T a b l e~ 1.
The values of the resonance width for different models. Model A corresponds to 
the  $SU(2)_C$ pure gluodynamics of Ref. \cite{LQCD:1}.
Model B describes the  $SU(3)_C$ LQCD with 2 quark flavors \cite{LQCD:2} and Model C 
is the $SU(3)_C$ LQCD with 3 quark flavors \cite{LQCD:3}. 
} 
\tabcolsep=7.5pt

\noindent
\begin{tabular}{ccccc}
\hline
 Model &     $\frac{90 \sigma}{\pi^2}$  & \quad $T_c$  &  $  \quad \Gamma_R (V_0, 0) $   &   \quad $ \Gamma_R (V_0, T_H) $ \\ 
Ref. & d.o.f.  &  \quad (MeV) & (MeV)  & (MeV) \\ \hline
 & & &  & \\
A  &  6  &  170     &    410 &   1420  \\
& & &  & \\
A &   6 &  200  
&   616 &   2133  \\
 & & &  & \\ \hline
 & & &  & \\
B  & 37  &  170 &   391   &  1355  \\
 & & &  & \\
B  & 37  &  200 &   587   &    2034  \\ 
 & & &  & \\ \hline
 & & &  & \\
C  & $\frac{95}{2} $ &  196 &   596    &  2066  \\
 & & &  & \\
 \hline 
\end{tabular}

}

\vskip8mm

\section{Concluding remarks}
Here we discussed   the  novel statistical approach to study  the QGP bags with  the  medium dependent finite width. 
This approach   is based on the Hagedorn-like mass spectrum of bags modified by the surface free energy of bags and by the bag  width. 
We found   that the volume dependent width of the QGP bags $\Gamma (v) = \gamma\, v^\frac{1}{2}$ 
leads to the Hagedorn mass spectrum of free heavy/large  bags.  Such a behavior of a width
also allows one  to explain a substantial  deficit of heavy hadronic resonances in the experimentally observed  mass spectrum and to resolve  the second conceptual problem introduced earlier. 

Further we considered the case of high temperatures and   derived  the general form   for  the bag pressure $p^+$ which accounts for the effect of finite  width in the EoS. We showed  that the obtained spectrum 
itself  cannot  explain the absence of directly observable QGP bags in the high energy nuclear 
and elementary particle collisions.

Then we studied the case  $T \le  c_\pm T_H$ (with $c_\pm <1$)   and found  the novel physical effect, the {\it subthreshold suppression} of heavy and large QGP bags. 
Such an effect  occurs due to the fact 
that at low  $T$ the most probable mass of heavy bags $\langle m \rangle$  is negative
and, thus, is below the lower cut-off  $M_0$ of
the continuous mass spectrum. 
Hence only the lightest  bags of mass about $M_0$ and of  smallest volume $V_0$
may contribute into the resulting spectrum, but such QGP bags will be indistinguishable  
from the low-lying  hadronic resonances.  Thus, the FWM resolves the first conceptual problem
which 
we discussed in the Introduction. 

Also we showed how  the FWM is able to reproduce a few  EoS of the  QGP  and 
demonstrated that the low $T$ pressure $p^-$ naturally  reproduces the linear $T$ dependence of   the LQCD  pressure for nonzero width coefficient.  Note that the linear $T$-dependence of the LQCD pressure  evidences for its  nonperturbative  origin. Therefore,  the ability of the FWM to reproduce  such a result is itself  a strong argument that this model catches the correct physics. 

Moreover, the derivation of two different regimes  $\langle m \rangle > 0$ and $\langle m \rangle \le 0$
and two corresponding dependences of the QGP pressure on the width  $\Gamma (v)$ given by Eqs. (\ref{PposM}) and  (\ref{pnegM}), respectively,  led us to an important conclusion that 
the consistent  and noncontradictory  statistical description of  an infinite bag  can be achieved, if and  only  if   $\Gamma(v) = \Gamma_1(v)$.  As one can see from (\ref{PposM}) and  (\ref{pnegM}) only such a choice  of the volume dependent  width   ensures  an  existence of  pressure for
an infinite QGP bag. 

A detailed study of the QGP EoS allowed us to  approximately  estimate  the volume dependent width  from the fresh LQCD data, which were analyzed in this work. 
As we showed these estimates are not sensitive to the number of elementary degrees of freedom of the LQCD, while they are sensitive to the transition temperature value.  
Our estimates of the volume dependent width look
very  promising  for heavy ion phenomenology since they introduce 
the new time scale into  play. 
A detailed discussion of  the Regge trajectories of the FWM  QGP bags   along with some possible experimental consequences  can be   found in  \cite{Reggeons:08}
and  \cite{FWM:08}, respectively.



\vspace*{0.4cm}

We are thankful  to P. Braun-Munzinger, H. Gutbrot,  D. H.  Rischke,
J. Stachel, H. St\"ocker and D. Voskresensky  for fruitful discussions and important comments.
One of us, K.A.B., thanks the department KP1 of GSI, Darmstadt,  for a warm hospitality.  
The authors express their gratitude to  T. V. Ivanchuk for  the   help in preparing the Ukrainian version of the manuscript. 
The research made in this paper  
was supported in part   by the Program ``Fundamental Properties of Physical Systems 
under Extreme Conditions''  of the Bureau of the Section of Physics and Astronomy  of
the National Academy of Science of Ukraine.



\end{document}